\documentclass[aps,twocolumn,nofootinbib,groupedaddress,amsfonts,floatfix]{revtex4-1} 
\usepackage{graphicx,amsmath,amssymb,amstext}
\usepackage{amssymb,amsbsy,amsfonts,amsthm,color}

\usepackage{epsfig}
\usepackage{graphicx}
\usepackage{subfigure}

\graphicspath{{Figures/}}

\usepackage{color}
\newcommand{\Beq}{\begin{eqnarray}}
\newcommand{\Eeq}{\end{eqnarray}}

\newcommand{\eqn}[1]{Eqn. (\ref{#1})}
\newcommand{\nn}{\nonumber \\}

\def\lsim{\mathrel {\vcenter {\baselineskip 0pt \kern 0pt \hbox{$<$} \kern 0pt \hbox{$\sim$} }}}

\def\gsim{\mathrel {\vcenter {\baselineskip 0pt \kern 0pt \hbox{$>$} \kern 0pt \hbox{$\sim$} }}}

\newcommand{\twodsquaremat}[4]{\left( \begin{array}{cc}#1 & #2\\ #3 & #4 \end{array} \right)}

\newcommand{\Tr}{\mathrm{Tr}}
\newcommand{\SEnv}{S {\cal E}}
\newcommand{\SE}{{\cal E}}

\begin{document}

\title{The Quantum Information of Cosmological Correlations}
\date{\today}

\author{Eugene A. Lim${}$}
\email{eugene.a.lim@gmail.com}

\affiliation{${}$Theoretical Particle Physics and Cosmology Group, Physics Department,
Kings College London, Strand, London WC2R 2LS, United Kingdom}

\begin{abstract}
It has been shown that the primordial perturbations sourced by inflation are driven to classicality by unitary evolution alone. However, their coupling with the environment such as photons and subsequent decoherence renders the cosmological correlations quantum, losing primordial information in the process. We argue that the quantumness of the resulting cosmological correlations is given by \emph{quantum discord}, which captures non-classical behavior beyond quantum entanglement. By considering the environment as a quantum channel in which primordial information contained in the perturbations is transmitted to us, we can then ask how much of this information is inaccessible. We show that this amount of information is given by the discord of the joint primordial perturbations-environment system. To illustrate these points, we model the joint system as a mixed bi-modal Gaussian state, and show that quantum discord is dependent on the basis which decoherence occurs.
\end{abstract}

\maketitle
\section{Introduction}

The paradigmatic theory of the early universe is inflation \cite{Guth:1980zm}, which posits that an early accelerated expansion of space drove hitherto causally correlated spacetime events outside the Hubble Horizon, laying down super Hubble horizon size correlations.  This  sets the stage for observers like us undergoing more prosaic expansion in the late universe to make the observation that the Universe is highly homogenous and isotropic. Crucially, in inflation, initial seeds of cosmological perturbations are sourced by fluctuations of the quantum vacuum of the inflaton field which drives inflation \cite{Guth:1982ec}. These primordial fluctuations are the progenitors of the density perturbations -- stars, galaxies, planets -- that we observed in the present Universe. This occurs as the inflaton decays into standard model particles (such as photons) and then interact with gravitational forces in a fairly complicated but straightforward manner to generate hot and cold regions in space. These manifest themselves as slightly hot and slightly cold (to one part in $10^5$) spots on the Cosmic Microwave Background (CMB). These temperature anisotropies were spectacularly confirmed in the past ten years \cite{Ade:2013uln}. 

To be specific, we observed the angular power spectrum $P_l$ of the temperature anisotropies $\Delta T/T({\bf n})$ in the direction ${\bf n}$  decomposed into spherical harmonics $a_{lm}$ 
\begin{equation}
P_{l} = \langle a_{lm} a_{lm}\rangle~,~\frac{\Delta T}{T_{\mathrm{CMB}}}({\bf n}) = \sum_{l,m}Y_{l,m}({\bf n})a_{lm} \label{eqn:powerspectrum}
\end{equation}
where  angled brackets means averaged over all $m$. Hence, our observations of the CMB is encoded as a set of $l$-valued classical probability distribution functions (pdf) $P_l$ with an ensemble of $m$ observations per $l$. Inflation predicts that $P_l$ to be very nearly scale invariant and highly Gaussian, both predictions which were confirmed to a very high precision \cite{Ade:2013uln}. In contrast to classical pdfs, quantum states are described by density matrices, which presumably might contain some valuable (to a physicist) quantum information. Given our measurements on the CMB, we can then ask the question \emph{are the cosmological correlations that we measure still possess quantum properties}?


To answer this question requires us to understand the structure of the quantum state of not just the cosmological correlations themselves, but also its decoherence by interactions with the photons and matter fields environment which ultimately serve as our detectors of these primordial fluctuations \cite{Polarski:1995jg}. In previous studies of this process, it is shown that the primordial perturbations are driven into a highly squeezed and classical state via unitary evolution during inflation \cite{Grishchuk:1990bj,Kiefer:2006je,Kiefer:1999sj,Polarski:1995jg,Albrecht:1992kf}. In this paper, we confirm the results of these studies that the primordial perturbations are indeed classical in the sense that the phase information of the perturbations are driven to very tiny values\footnote{Operationally, the ``decaying'' mode of the inflationary perturbations gets driven to very small values, carrying the phase information with it.}.  Furthermore, we prove that this classicalness is measurement basis independent, as long as we limit our consideration to the closed system of primordial perturbations.

However, in real life, we do not have direct access to these primordial perturbations. Instead, we observe things like photons and the matter distribution in the Universe -- things which have long decohered and become classical in the sense that they are resilient to further monitoring by our clumsy classical instruments. Primordial perturbations couple to this environment, which through some process of decoherence picks up a basis of which we make measurements on. We subsequently observe a small subset of this environment to learn about the primordial information. Furthermore, even though the primordial perturbations fresh out of inflation may be classical, the decoherence process (ironically) generates couplings to the environment, rendering the cosmological correlations we observe quantum. In other words, the environment serves as a quantum channel in which the classical information of the primordial perturbations is transmitted to us.   We can now ask, \emph{how much of the primordial quantum information can we recover from our late time observations}? A possible candidate answer is to calculate the entanglement entropy of the primordial perturbations immersed in an environment \cite{Kiefer:2006je}. 

However, it is a well known fact that mixed quantum states, such as that of the joint primordial perturbations-environment system, has no unique entanglement measure \cite{RevModPhys.81.865}. More importantly, while other entropy measures have been calculated for cosmological correlations \cite{Campo:2008ij,Campo:2008ju}, it is now known that the quantum nature of any system goes beyond its entanglement measure, and a robust measurement of ``quantumness'' and information loss instead is \emph{quantum discord} \cite{Ollivier:2001aa,Henderson:2001aa}. In particular, quantum discord captures quantum correlations beyond entanglement of mixed state density matrices. Furthermore, we will show that the quantum discord also quantifies the amount of  information of the primordial perturbations \emph{inaccessible} via observations of the environment.

In this paper, we compute the quantum discord for a model of this joint system, and show that the amount of quantumness and information loss of this joint system is dependent on both the physics of decoherence and the basis it ``chose''. In particular, for the latter, the question of what is the pointer basis (of the primordial states) that the environment picks up is unclear. One possibility, championed by Kiefer-Polarski-Starobinsky \cite{Kiefer:2006je}, is that the position basis of the inflaton field is the natural pointer basis of decoherence.  On the other hand, Campo and Parentani suggest that the coherent basis is the natural basis \cite{Campo:2005sv}. In this paper, we do not attempt to consider all these possibilities, but instead we will construct a model which shows the basis dependence explicitly. One can also try to construct a decoherence model -- for example Ref. \cite{Burgess:2006jn,Burgess:2014eoa} propose that decoherence occurs via interactions of the high and low frequency modes. 

The paper is organized as follows. In Section \ref{sect:Discord}, we review the notion of quantum discord. In Section \ref{sect:gaussianmodel}, we derive the primordial perturbations from inflation and construct a phenomenological model joint primordial perturbations-environment state. In Section \ref{sect:cosmoDiscord}, we compute the quantum discord of this joint state, and show that it is dependence on basis of decoherence/measurement and argue that it also quantifies the inaccessible information. We conclude in Section \ref{sect:conclusions}.



\section{Quantum Discord \label{sect:Discord}}

In this section, we describe quantum discord. The impatient can skip right ahead to \eqn{eqn:finaldiscordequation}.

Given a pure bipartite system $AB$ described by the density matrix $\rho_{AB}$, the unique measure of quantum correlations is the entanglement entropy $S(\rho_A)=S(\rho_B)$ where $S$ is the Von Neumann entropy
\begin{equation}
S(\rho) = -\Tr \rho \log \rho.
\end{equation}
and the reduced states $\rho_A = \Tr_B (\rho_{AB})$ and $\rho_B = \Tr_A(\rho_{AB})$ where $\Tr_X$ means a partial trace over the $X$ space.

However, for mixed bipartite systems, there is no unique entanglement measure. Nevertheless, there is a notion of no-entanglement -- a mixed $AB$ state is not entangled if it can be written in the separable form
\begin{equation}
\rho_{AB} = \sum_i p_i|a_i\rangle\langle a_i|\otimes |b_i\rangle\langle b_i|
\end{equation}
where $\{|a_i\rangle\}$ and $\{|b_i\rangle\}$ are some complete and orthogonal basis for the separate systems $A$ and $B$ respectively.

Ollivier and Zurek \cite{Ollivier:2001aa}, and independently Henderson and Vedral \cite{Henderson:2001aa} proposed a new measure, \emph{quantum discord}, as a more robust measure of quantum correlations (for a recent review, see \cite{Modi:2012aa}).  The idea is the following. Consider the usual classical information provided by the Shannon entropy
\begin{equation}
H(A) = -\sum_a p(A=a) \log p(A=a),
\end{equation}
where we have summed over all possible realizations of the pdf $p(A)$.

A bipartite classical system $AB$ with some overlap $A\cap B$ can be  described by the joint pdf $p(A,B)$ where $A$ and $B$ are the parameters. This overlap means that if we know something about the state of $B$, then we will learn something about $A$. This ``something'' is quantified by the \emph{classical mutual information} which is given by
\begin{equation}
J(A:B) = H(A)-H(A|B) \label{eqn:classicalJ}
\end{equation}
where the conditional entropy $H(A|B) = \sum_b p(B=b)H(A|B=b)$, and the separate marginalized pdfs $p(A)$ and $p(B)$ are derived from the joint pdf $p(A,B)$ via
\begin{equation}
p(A) = \sum_b p(A,B=b)~,~p(B)=\sum_a p(A=a,B)
\end{equation}
and $H(A|B=b)$ is the information contained in the posterior pdf (i.e. information contained in $A$ knowing $B=b$). Classical joint pdfs obey Bayes' Theorem
\begin{eqnarray}
p(A|B=b) &=& \frac{p(A,B=b)}{p(B=b)}, \nn
H(A|B=b) &=& -\sum_a p(A=a|B=b) \log p(A=a|B=b), \nn
&&
\end{eqnarray}
which we can plug into \eqn{eqn:classicalJ} to obtain another measure of classical information
\begin{equation}
I(A:B) = H(A)+H(B)-H(A,B).
\end{equation}
Classically, $I(A:B)$ and $J(A:B)$ are completely equivalent, so $I(A:B)-J(A:B)=0$. However, quantum mechanically, the notion of the posterior pdf $H(A|B)$ is ill-posed -- suppose $A$ and $B$ are correlated in a quantum manner, a measurement of $B$ which we need to know to compute $H(A|B)$ may change the state of $A$! In other words, Bayes' theorem do not translate through quantum correlations (since quantum correlations cannot be expressed as joint pdfs). The main idea is to use the difference in the quantum version of $I$ and $J$ to define quantum discord as a measure of quantum correlations.

For quantum correlations, the analogous quantity to the Shannon entropy is the Von Neumann entropy\footnote{For mixed states, this is not a unique choice since there is no unique measure of entanglement. See \cite{RevModPhys.81.865} for a thorough review.} so we can replace
\begin{equation}
H(A) \rightarrow S(A) = -\Tr \rho_A \log \rho_A
\end{equation}
where $\rho_A$ is now a density matrix. The quantum generalization of $I(A:B)$ is straightforward
\begin{equation}
{\cal I}(A:B) = S(A)+S(B)-S(A,B).
\end{equation}
On the other hand, the quantum version of ${\cal J}(A:B)$ requires the conditional entropy $S(A|B)$. Since a measurement of $B$ will necessarily affect $A$ in a quantum state, one can \emph{define} a notion of conditional entropy by first considering a set of projection operators  $\{\Pi_k^B\}$ on $B$ which forms a positive operator valued measure (POVM) obeying the partition of unity $\sum_k \Pi_k^B = {\bf 1}$. POVMs are generalization of complete sets in that they are not necessarily orthogonal -- we will use these POVMs as probe states on $B$.

Now given a state $\rho_{AB}$, a measurement of the state associated with $\Pi_k^B$  transforms the state into
\begin{equation}
\rho_{AB} \rightarrow \frac{\rho_{AB}\Pi_k^B}{P_k}
\end{equation}
with probability $P_k = \Tr_{AB}(\rho_{AB}\Pi_k^B)$. Given access only to $A$, we describe the state by tracing over $B$, i.e.
\begin{equation}
\rho_{A|B=\Pi_k^B} \equiv \Tr_B \frac{\rho_{AB}\Pi_k^B}{P_k}. \label{eqn:conditionedoutcome}
\end{equation}

Then if one makes a set of all possible measurements $\{\Pi_k^B\}$, the conditional entropy is defined to be 
\begin{equation}
S(A|B=\{\Pi_k^B\}) \equiv \sum_k P_k S(\rho_{A|B=\Pi_k^B})
\end{equation}
so we can construct the quantum version of $J(A:B)$
\begin{equation}
{\cal J}(A:B)_{\Pi_k^B} = S(A)-S(A|B=\{\Pi_k^B\}) \label{eqn:JHolevo}
\end{equation}
where the subscript on the LHS is to remind us that ${\cal J}$ is dependent on the choice of POVMs. As we will see later, ${\cal J}(A:B)_{\Pi_k^B}$ is actually the Holevo information from $A$ to $B$ \cite{Zwolak:2013aa} -- this fact will be important when we consider the maximum possible information we can recover from the primordial perturbations.

The discord is defined to be the difference between ${\cal I}$ and ${\cal J}$
\begin{eqnarray}
\delta(A:B)_{\Pi_k^B}  &=& {\cal I}(A:B)-{\cal J}(A:B)_{\Pi_k^B} \nn
&=& S(B)-S(A,B)+ \sum_k P_k S(\rho_{A|B=\Pi_k^B}) \geq0. \nn
&& \label{eqn:finaldiscordequation}
\end{eqnarray}

It is clear that if discord vanishes, then the state is truly classical since it can be represented by joint pdfs. Note that discord is non-symmetric under the interchange of $A \leftrightarrow B$  as the conditional entropy is non-symmetric. 

Crucially, \emph{mixed separable states can have nonzero discord} -- discord captures non-classical correlations beyond entanglement. In general, discord is also a function of the probe states $\{\Pi_k^B\}$ and hence is measurement basis dependent. One can make it measurement basis independent by insisting that it is minimized over the field of all possible sets of probe states $\Pi_{k,q}^B$ labeled by $q$ 
\begin{equation}
\delta(A:B)\equiv \underset{q}{\mathrm{inf}}(\delta(A:B)_{\Pi_{k,q}^B}).
\end{equation}
This ``measure independent discord'' is  invariant under local unitary transforms, i.e. the discord for  $\rho_{AB}$ is the same as that of $(U_A\otimes U_B) \rho_{AB} (U_A \otimes U_B)^{\dagger}$ where $U_A$ and $U_B$ are local unitary transforms on $A$ and $B$ respectively. However, since we are considering cosmological correlations, we are not allowed to willy-nilly rotate the entire universe, so we will use the measure dependent discord \eqn{eqn:finaldiscordequation} throughout the paper.

\section{Gaussian model of joint perturbations-environment state} \label{sect:gaussianmodel}

During inflation, the initial conditions of the perturbation modes are sourced by quantum fluctuations of its vacuum, the so-called Bunch-Davies vacuum (for a thorough review, see \cite{Mukhanov:1990me}). In linear theory, one can work in Fourier space and label the modes by its co-moving momentum $k$.  Following \cite{Grishchuk:1990bj,Polarski:1995jg,Albrecht:1992kf}, we work in the Schr{\"o}dinger's picture. If we define the perturbations as $y_k$, the Hamiltonian can be written as 
\begin{equation}
\hat{H}_k =\frac{1}{2}\left(p_k^2+k^2y_k^2+\frac{2a'}{a'}y_kp_k\right) \label{eqn:Hamiltonian}
\end{equation}
where the canonical momentum $p_k =\partial{L(y,y')}/{\partial y'} = y'-a'/ay$ where $a(\eta)$ is the cosmic scale factor,  and $\eta$ is the co-moving time. Primes denote derivatives with respect to comoving time $\eta$.   Using the  Schr{\"o}dinger's Equation 
\begin{equation}
i\hbar \frac{\partial \psi(y,\eta)}{\partial \eta}=\hat{H}_k \psi(y,\eta) \label{eqn:SchrEqn}
\end{equation}
we can evolve the wave function $\psi(y_k,\eta)$. The solution to \eqn{eqn:SchrEqn} is
\begin{equation}
\psi(y,\eta)=\left(\frac{2\Omega_R(\eta)}{\pi}\right)^{1/4}\exp(-(\Omega_R+i\Omega_I)y^2),
\end{equation}
where $k^{-1} \Omega_R=(\cosh 2r + \cos 2\varphi \sinh 2r)^{-1}$ and $\Omega_I=-\Omega_R\sin 2\varphi \sinh 2r$ encode the dynamics of the background $a(\eta)$. Inflation is a phase of near de Sitter space, which in this limit we have
\begin{equation}
\sinh 2r = \frac{a H_I}{2k}~,~\cos 2\varphi = \tanh r
\end{equation}
where the number of e-folds of inflation is given by 
\begin{equation}
r=\log \frac{a}{a_i} > 60,
\end{equation}
which also happens to coincide with the \emph{squeezing parameter}. At the end of inflation, $\Omega_R \rightarrow ke^{-2r}~,~\Omega_I \rightarrow -ke^{-r}$. Ignoring decoherence for the moment, we can write down the density matrix for the cosmological perturbation pure state
\begin{eqnarray}
\rho_S(y,y') &= &\psi(y,\eta)^* \psi(y',\eta) \nn
&=& \frac{2\Omega_R}{\pi}e^{-[\frac{\Omega_R}{2}(y-y')^2-\frac{\Omega_R}{2}(y+y')^2-i\Omega_I(y^2-y'{}^2)]}.\nn
&&\label{eqn:rhoseparable}
\end{eqnarray}
In general $y$ is complex, and contain information on both the positive and negative frequency $\pm{\bf k}$ modes. One way to think about this is that the positive and negative frequency sectors jointly form an entangled bipartite system when they are produced -- just as in Hawking radiation\footnote{I am grateful to Lam Hui and Riccardo Penco for pointing this out to me. See \cite{Campo:2005sy}.}. In perturbation theory, this entanglement is tiny (3rd order at the most), and we can ignore their couplings and consider only the real part of $y$. At late times, then $\Omega_R \gg 1$ as $r\gg 1$, the off-diagonal terms get suppressed and we obtain a density matrix which is highly diagonal -- phase information is driven to very small values. Suppose now that we have an inflaton detector, and can directly make observations in the $y$-basis, then what we obtain is simply a classical pdf in the $y$-basis. The off-diagonal phase information have been ``lost''\footnote{Of course, evolution here is still unitary, so one can in principle turn the clock backwards and recover the phase.}, so the system no longer looks ``quantum'' in the sense that there can be no interference effects. Note that this loss of coherence is qualitatively different from the quantum effects of entanglement (since there is nothing to entangle with in a single partite state).

We now consider entanglement with the environment. Kiefer, Polarski and Starobinsky \cite{Kiefer:2006je} following \cite{Joos:1984uk} argue that macroscopic localization of the inflaton particles (via a large number of scattering processes with the environment), leads to the following \emph{ansatz} for the density matrix
\begin{equation}
\rho_S(y,y') \rightarrow \rho_S'(y,y') = \rho_S(y,y') \times \exp\left[-\frac{\zeta}{2}(y-y')^2\right] \label{eqn:ansatz}
\end{equation}
where the ``decoherence parameter'' $\zeta$ encodes the strength and intensity of the interactions and obeys
\begin{equation}
\Omega_I \gg \zeta \gg \Omega_R. \label{eqn:zeta}
\end{equation}
The first inequality is required to preserve the fact that the pointer basis $\{y_k\}$ is still squeezed, while the 2nd inequality is chosen such that the interactions with the environment ``blurs'' the sharpness of the original squeezed state. 

\begin{figure}
\begin{center}
\includegraphics[width=9cm]{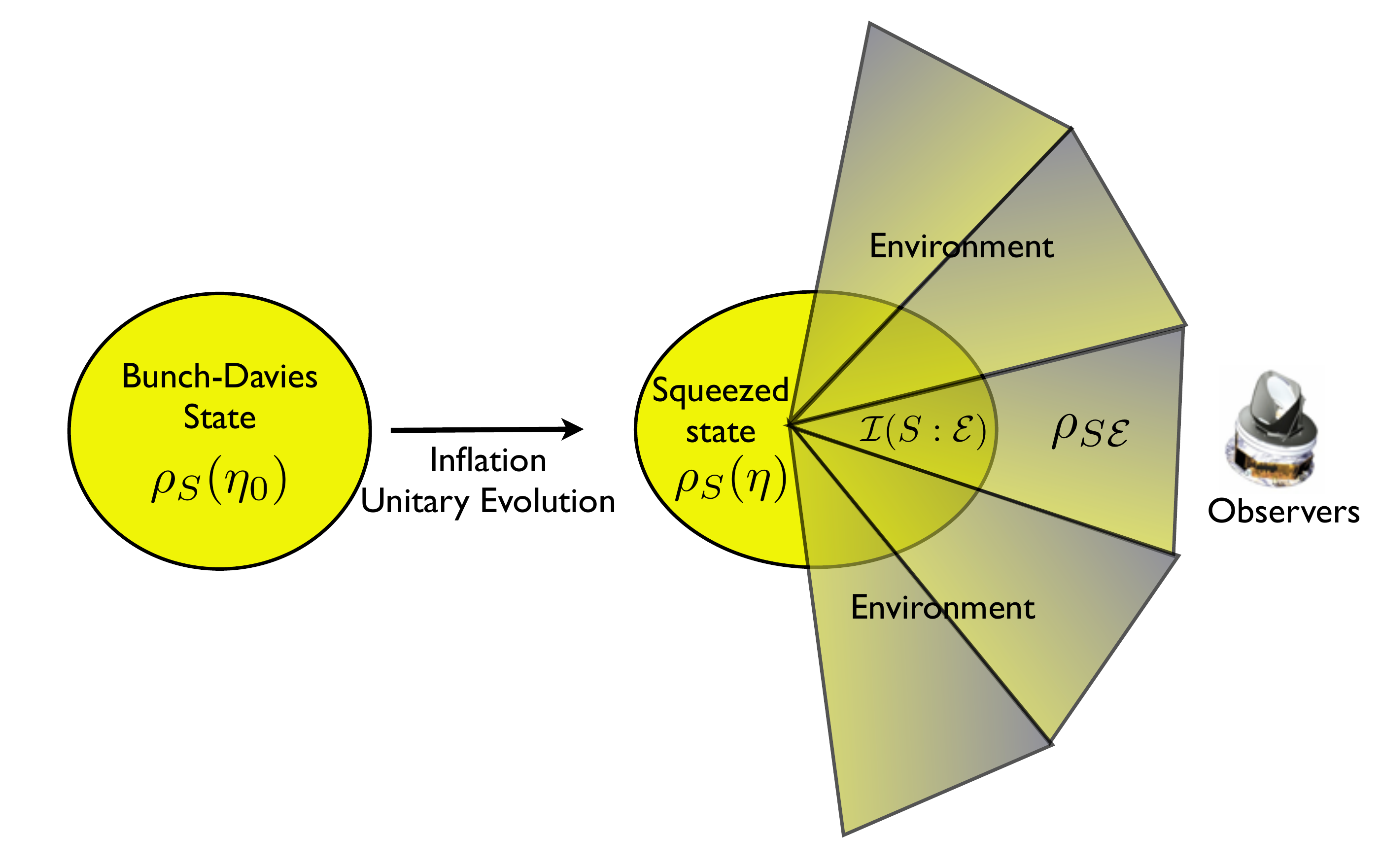}
\caption{Primordial perturbations start off in the Bunch-Davies vacuum state $\rho(\eta_0)$, evolve via unitary evolution during inflation into the squeezed state $\rho_S(\eta)$. These perturbations source the fluctuations in the Cosmic Microwave Background which seeds the formation of large scale structure of galaxies,  collectively forming the environment in which we can observe, decohering in the process. The environment hence acts as a quantum channel which transmits the primordial information to us.  We make (classical) observations on a small subset (a single triangle) of the environment to infer the nature of the primordial fluctuation.}
\label{fig:B}
\end{center}
\vskip -20pt
\end{figure}

These couplings generically  lead to a loss of quantum information of the original pure state via leakage into the environment, i.e. $\rho_S'$ is a mixed state. Our cosmological observations do not directly probe the perturbations, instead we probe the environment for which it couples to -- photons and matter fields -- in order to learn about the primordial perturbations. Put in another way, the environment acts as a quantum channel of which the original primordial perturbations is transmitted (see Figure \ref{fig:B}). Our observations are classical of course (in the sense that our data is encoded as a set of classical pdfs).  As it is well known, the Holevo information \cite{Holevo:1973a} puts an upper limit on how much of this primordial quantum information can be transmitted to us through this channel. We will also compute this loss of information in the following.

Now it is not surprising that this loss is highly dependent on the exact form of the interactions between the primordial perturbations and the environment. In principle, one should be able to construct a model of decoherence by considering all possible couplings, both direct and gravitational, between the primordial perturbations $S$ and the environment ${\cal E}$. We emphasise that ${\cal E}$ refers to the environment that we can make observations off -- it is a very small subset of the entire universe. To construct this joint state from first principles is highly complicated. In this work, we will instead take a phenomenological tack by constructing a joint density matrix $\rho_{\SEnv}$ such that
\begin{equation}
\Tr_{{\cal E}} (\rho_{\SEnv}) = \rho_S'. \label{eqn:jointtrace}
\end{equation}
In general, there is no unique $\rho_{\SEnv}$. Fortunately, it turns out that the reduced state \eqn{eqn:rhoseparable} is a single mode Gaussian quantum state which allows us to make progress (for a recent review on Gaussian quantum states, see \cite{Weedbrook:2012aa}). Given any Gaussian quantum state, it is possible to construct a purification of $\rho_S'$, i.e. a pure two mode Gaussian state $\rho_{\SEnv}$  \cite{PhysRevA.63.032312} which obeys \eqn{eqn:jointtrace}. We can then parameterize around this pure two mode Gaussian to construct a generic mixed joint state. 

To do this, instead of fiddling around with density matrices, it is convenient to work in the equivalent Wigner distribution picture which we can obtain by Weyl transforming the density matrix \eqn{eqn:rhoseparable} via
\begin{eqnarray}
W(y,p)& =& \frac{1}{\pi}\int_{-\infty}^{\infty} dx~ e^{2ipx}\rho(y-x/2,y+x/2), \nn
&=& \frac{1}{\pi}\exp\left[-\frac{2(p+\Omega_Iy)^2}{\Omega_R+\zeta}+2\Omega_R y^2\right].
\end{eqnarray}
We can further transform \eqn{eqn:rhoseparable} into the canonical Gaussian form with $y\rightarrow y/2\sqrt{k}$, $p\rightarrow (1/2)\sqrt{k}p$, $\alpha\equiv -\Omega_I/k=e^{-r}$, $\lambda \equiv \Omega_R/k= e^{-2r}$ and $\xi = \zeta/k$, such that $\alpha>0, \lambda >0$. Then the Wigner distribution becomes
\begin{eqnarray}
W(y,p)&=&\frac{1}{\pi}\sqrt{\frac{\lambda}{\lambda+\zeta}}\exp\left[-\frac{1}{2}\left[\frac{(p-\alpha y)^2}{\lambda+\xi}+\lambda y^2\right]\right] \nn
&=& \frac{1}{\pi}\sqrt{\frac{\lambda}{\lambda+\zeta}}\exp\left[-\frac{1}{2}{\bf y}\sigma_S^{-1}{\bf y}^T\right]  \label{eqn:GaussianWigner}
\end{eqnarray}
where ${\bf y} = \{y,p\}$ and the \emph{covariance matrix} $\sigma_S$ of the single mode Gaussian state is given by
\begin{equation}
\sigma_S = \twodsquaremat{\frac{1}{\lambda}}{\frac{\alpha}{\lambda}}{\frac{\alpha}{\lambda}}{\lambda+\xi+\frac{\alpha^2}{\lambda}}. \label{eqn:CMSstate}
\end{equation}
The state is normalized to
\begin{equation}
\int_{-\infty}^{\infty} {\cal D}{\bf y}~W_S = 4\pi.
\end{equation}

A general $N$ mode Gaussian state describes dynamics of a $2N$ vector $\{y_1,p_1,y_2,p_2,\dots,y_N,p_N\}$. The conditions on $\sigma$ for this to be a physical state is $\sigma+i\Omega>0$ (which is derived from imposing pair wise canonical commutation relations on $x_i$ and $p_i$)  where $\Omega$ is the symplectic form given by
\begin{equation}
\Omega = \omega\oplus \omega\oplus\omega \oplus \dots \omega~N~\mathrm{times}~,~\omega = \twodsquaremat{0}{1}{-1}{0}. \label{eqn:omega}
\end{equation}

We can then construct a two parameter $(\beta,\tau)$ mixed 2-mode Gaussian state with ${\bf y} = \{y_1,p_1,y_2,p_2\}$ whose convariance matrix $\sigma_{\SEnv}$ is given by (see Appendix \ref{app:purification} for the derivation)
\begin{widetext}
\begin{equation}
\sigma_{\SEnv} = \left(\begin{array}{cccc}
\frac{1}{\lambda} & \frac{\alpha}{\lambda} & \sqrt{\frac{\xi}{\lambda}}\left(1+\frac{\xi}{\lambda}\right)^{-1/4} & 0 \\
\frac{\alpha}{\lambda} & \xi+\frac{\alpha^2}{\lambda}+\lambda & 
\alpha \sqrt{\frac{\xi}{\lambda}}\left(1+\frac{\xi}{\lambda}\right)^{-1/4}
& -\sqrt{-1+\left(1+\frac{\xi}{\lambda} \right)\tau^2}(\lambda(\xi+\lambda))^{1/4} \\
\sqrt{\frac{\xi}{\lambda}}\left(1+\frac{\xi}{\lambda}\right)^{-1/4} & \alpha\sqrt{\frac{\xi}{\lambda}}\left(1+\frac{\xi}{\lambda}\right)^{-1/4} & \beta\sqrt{1+\frac{\xi}{\lambda}} & 0 \\
0 & -\sqrt{-1+\left(1+\frac{\xi}{\lambda} \right)\tau^2}(\lambda(\xi+\lambda))^{1/4} & 0 & \beta\sqrt{1+\frac{\xi}{\lambda}}
\end{array} \right) \label{eqn:mixedstategaussian},
\end{equation}
so its normalized Wigner distribution is
\begin{equation}
W_{\SEnv} =  \frac{1}{\pi}\frac{\lambda}{\sqrt{(-\xi+\beta(\xi+\lambda))(\lambda+(\xi+\lambda)(\beta-\tau^2))}}\exp\left[-\frac{1}{2}{\bf y}\sigma_{\SEnv}^{-1}{\bf y}^T \right]. \label{eqn:normWignerJoint}
\end{equation}
\end{widetext}
The state \eqn{eqn:mixedstategaussian} is a pure 2-mode Gaussian state when $\beta=\tau=1$. We hasten to add that while the joint state does not necessary have to be Gaussian, this assumption is  not unreasonable -- indeed it should be expected via the Central Limit Theorem as the environment itself is also highly Gaussian (i.e. our observations of the CMB has shown that it is highly Gaussian).  This is model is not the most general, but is chosen such that it illustrates the measurement basis dependence of discord.

The state \eqn{eqn:mixedstategaussian} can be rewritten in the following block form
\begin{equation}
\sigma_{\SEnv}=\twodsquaremat{{\cal A}}{{\cal C}}{{\cal C}^T}{{\cal B}}.
\end{equation}
where ${\cal A},{\cal B}$ and ${\cal C}$ are $2\times 2$ matrices, from which we can compute the following
\begin{eqnarray}
A& =& \det({\cal A})  = 1 + \frac{\xi}{\lambda}  \nn
B&=& \det({\cal B}) =  \beta^2\left(1 + \frac{\xi}{\lambda} \right) \nn
C&=& \det({\cal C}) = -\sqrt{\frac{\xi}{\lambda}\left[-1+\tau^2\left(1 + \frac{\xi}{\lambda} \right)\right]}  \nn
D&=& \det(\sigma_{\SEnv}) \nn
&=&\left[-\frac{\xi}{\lambda}+\left(1+\frac{\xi}{\lambda}\right)\beta\right]\left[1+\left(1+\frac{\xi}{\lambda}\right)(\beta-\tau^2)\right]. \nn
&&\label{eqn:ABCD}
\end{eqnarray}
Note that $C<0$, which implies that the state is entangled since $\rho_S'$ is a mixed state \cite{Weedbrook:2012aa}. In addition to the determinants, we can also calculate the symplectic eigenvalues $\nu_{\pm}$ which are particularly useful quantities of this matrix
\begin{equation}
\nu_{\pm}=\sqrt{\frac{1}{2}(\Delta\pm \sqrt{\Delta^2-4D})} \label{eqn:symplecticEV}
\end{equation}
where
\begin{eqnarray}
&\Delta \equiv A+B+2C \nn
&= \left(1+\frac{\xi}{\lambda}\right)(1+\beta^2)-2\sqrt{\frac{\xi}{\lambda}\left[-1+\left(1+\frac{\xi}{\lambda}\right)\tau^2\right]}.
\end{eqnarray}
The orientation angle $\alpha$ drops out of the calculation as expected since it is simply the phase of the primordial perturbations.

Physicality of the state imposes the conditions $A,B,\nu_{\pm} \geq 1$, with $\nu_{\pm}=1$ when the state is pure.  In the limit of $\xi/\lambda \gg 1$ as imposed by the second inequality of \eqn{eqn:zeta}, the first two conditions translate to
\begin{equation}
\beta^2,\tau^2\geq \left(1+\frac{\xi}{\lambda}\right)^{-1}.
\end{equation}
For simplicity, we impose the additional conditions $\beta \geq 1$ and $\tau^2 \gg (1+\xi/\lambda)^{-1}$, which leads to the condition $\tau^2 \leq \beta$.

\section{Quantum Discord of Cosmological Correlations} \label{sect:cosmoDiscord}

We can then proceed to compute the discord for the joint state \eqn{eqn:mixedstategaussian}. To do that, we need to choose a measurement basis. Again, by appealing to our observations, we choose generalized pure Gaussian POVM operators $\{\Pi^{\SE}\}$ as probe states \cite{Adesso:2010a}. Since they are Gaussian states, they can be described by the covariance matrix
\begin{equation}
\sigma_0 = R(\theta)
\twodsquaremat{\gamma}{0}{0}{\frac{1}{\gamma}}
R^T(\theta)
\end{equation}
where $R(\theta)$ is a rotation matrix, with $\theta$ being the ``alignment angle'' and $\gamma$ is the squeezing parameter. Note that the points $(\theta=\pi,\gamma)$ and $(\theta=\pi/2, 1/\gamma)$ are equivalent.  For each choice of $(\theta,\lambda)$, we can then generate \cite{Adesso:2010a} a set of single mode Gaussian probe states labeled by the complex parameter $z$, which forms a POVM with the partition of unity
\begin{equation}
\frac{1}{\pi}\int d^2z \Pi_{\SE}(z) = {\bf 1}.
\end{equation}

The discord can then be computed via \cite{PhysRevLett.105.020503,Adesso:2010a,Duan:2000aa}
\begin{equation}
\delta(S:{\cal E})_{\Pi_{\SE}(z)} = f(\sqrt{A})-f(\nu_-)-f(\nu_+)+f(\sqrt{\det \epsilon}) \label{eqn:discordGen}
\end{equation}
where
\begin{equation}
f(x) = \frac{x+1}{2}\log \left(\frac{x+1}{2}\right)-\frac{x-1}{2}\log \left(\frac{x-1}{2}\right),\label{eqn:feqn}
\end{equation}
and 
\begin{equation}
\epsilon = {\cal A}-{\cal C}({\cal B}+\sigma_0)^{-1}{\cal C}^{T},
\end{equation}
which is the covariance matrix of the state $\rho_{S|\Pi_{\SE}(z)}$ after the measurement $\Pi_{\SE}(z)$ (note that the state  is independent of $z$ \cite{PhysRevA.66.032316}).

\begin{figure}
\begin{center}
\includegraphics[width=5cm]{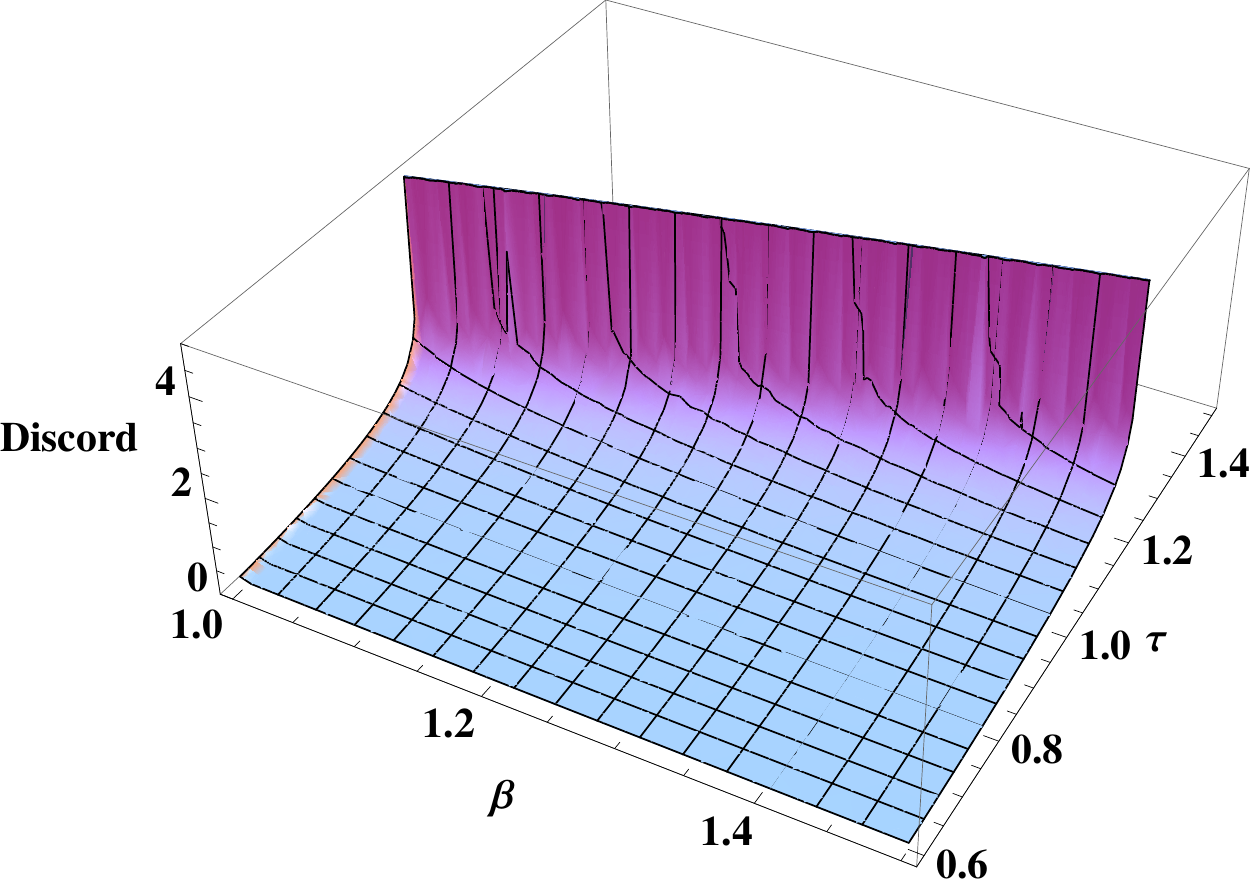}
\includegraphics[width=5cm]{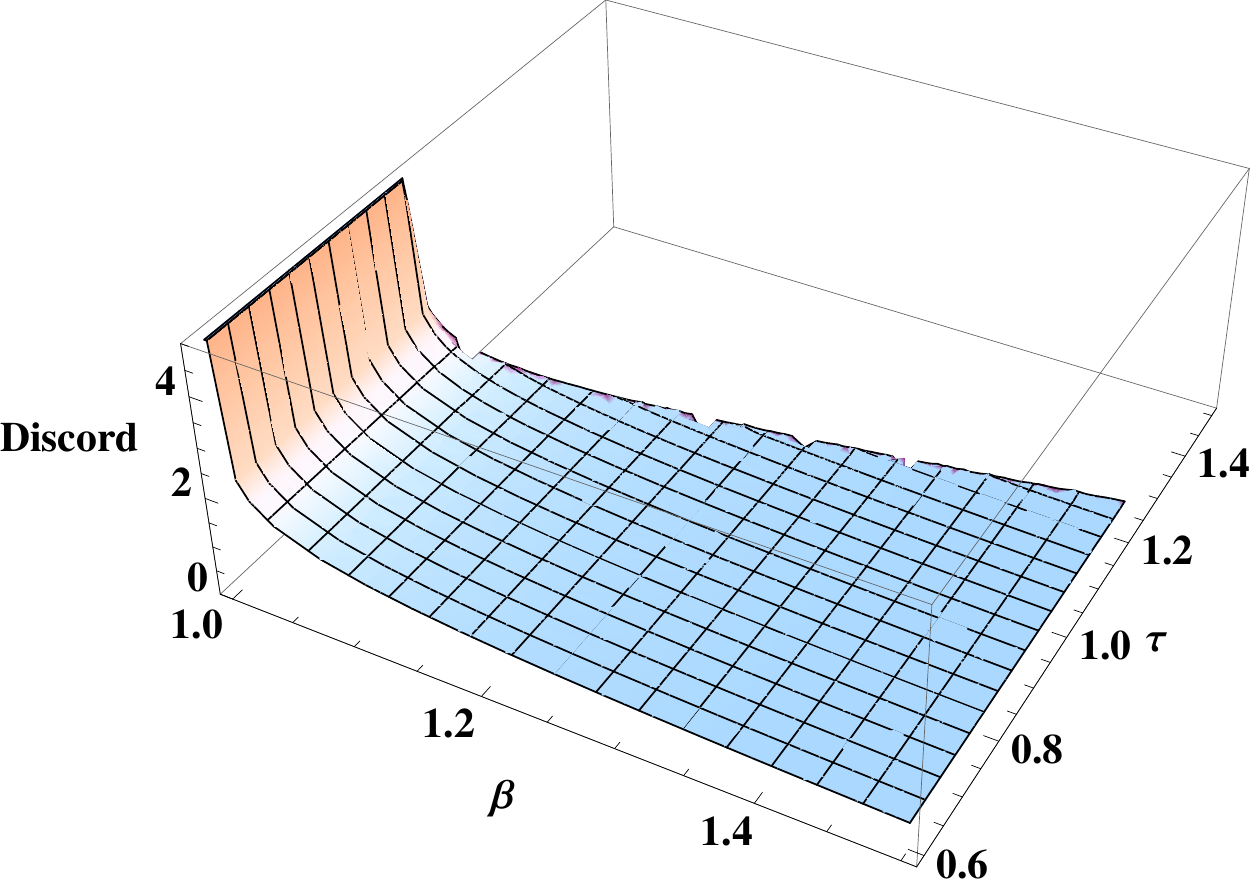}
\caption{The discord $\delta(S:{\cal E})$ of the joint perturbations-environment state $\rho_{\SEnv}$, parameterized by $(\beta,\tau)$. In this plot, we have chosen $\xi/\lambda= 10^5$, and the measurement basis $\gamma=10^5$ and $\theta=\pi/2$ (top) and $\theta=\pi$ (bottom). The maximum occurs at $\tau^2=\beta$ (top) and $\beta=1$ (bottom). The whiteout area indicates that the state is unphysical. 
\label{fig:A}}
\end{center}
\vskip -20pt
\end{figure}

The results of the calculation is shown in Figure \ref{fig:A}. The discord is highly dependent the choice of basis parameterized by $(\theta,\gamma)$. Maximum discord occurs at the points $(\theta=\pi/2,\beta=\tau^2)$ and $(\theta=\pi,\beta=1)$, and their equivalent points under the transformation $\gamma\rightarrow 1/\gamma$ and $\theta \rightarrow \theta+\pi/2$. We emphasise that despite the fact that these Gaussian states have positive definite Wigner functions, they are not classical as their discord is non-vanishing. In other words positivity of Wigner functions is an insufficient criteria for classical behavior -- Gaussian states have vanishing discord only in the case where they are products states \cite{PhysRevLett.105.020503}.

In the special case where $\sigma_{\SEnv}$ is pure (i.e. when $\beta=\tau=1$), $\nu_{\pm}=\det{\epsilon}=1$, and hence via \eqn{eqn:discordGen} we find (with $\chi \equiv \xi/\lambda$)
\begin{equation}
\delta(A:B)_{\Pi_{\SE}(z)} = f(1+\sqrt{1+\xi/\lambda}) \label{eqn:mindiscord1}
\end{equation}
which is also equal to the entanglement entropy of the perturbations traced over the environment $S(\Tr_{\cal E}\rho_{\SEnv})$ (first calculated in \cite{Kiefer:2006je}). This result is not surprising -- it is known that discord of a pure state is equivalent to its entanglement entropy \cite{Datta:2008aa}. Interestingly, if there is no decoherence $\xi=0$, the discord vanishes and the cosmological perturbations are indeed classical at the end of inflation, driven by unitary evolution alone. If all we have is direct access to the cosmological perturbations, and nothing else, then there will be no physical way we can distinguish them from a set of classical perturbations despite their quantum origins.  

Ironically, it is the decoherence of these perturbations via couplings with the environment, and hence the generation of genuine couplings (including true entanglement) between the perturbations and the environment which render the cosmological correlations quantum. The strength of these quantum correlations gained through this coupling depends on the strength of the couplings between the perturbations and the environment, and information is lost in this process. How much of the information is lost depends on the measurement basis (or more practically, the decoherence basis). Of course, even though the measured cosmological correlations are now quantum (and hence subject to tests of quantumness such as Bell's \cite{Campo:2005sv,Campo:2005sy} or the CHSH \cite{PhysRevLett.23.880} inequality), detection of any quantumness do not imply that the primordial correlations are quantum. 

On the other hand, one can ask the question \emph{how much information is lost during the decoherence process}. In other words, what is the maximum amount of information we can learn about primordial perturbations? For any given quantum channel described by $Q=\{P_k,\rho_k\}$, the upper bound of classical information transmitted is bounded by an upper limit known as the Holevo information $\chi(Q)$ \cite{Holevo:1973a}.  The Holevo bound $\chi(Q) \geq I(A:B)$ refers to maximum amount the \emph{classical} mutual information $I(A:B)$ accessible about system $A$ from system $B$, through quantum channel $Q$. 

The answer to our question is provided by a recent observation shown in Ref. \cite{Zwolak:2013aa}, whose authors noted that for a bipartite system $\SEnv$, the \emph{quantum} mutual information (the knowledge we learn about the perturbations $S$ given our measurements of  the environment ${\cal E}$) is a sum of quantum discord and the Holevo information $\chi(P_k, \rho_{S|\Pi_k^{\cal E}})$

\begin{equation}
{\cal I}(S:{\cal E}) = \delta(S:{\cal E})_{\Pi_k^{\cal E}} + \chi(P_k, \rho_{S|\Pi_k^{\cal E}}). \label{eqn:conservation}
\end{equation}
\eqn{eqn:conservation} means that the discord quantifies exactly how much of the information contain in the quantum correlations is \emph{inaccessible} to cosmologists -- i.e. information loss during the measurement. To see this, we first derive relation \eqn{eqn:conservation}.  Consider the definition of the Holevo information. Suppose we want to transmit some classical information encoded in the set of classical probabilities $\{P_k\}$ labeled by $k$, using a quantum channel. We can construct a quantum state made out of the ensemble $\{P_k,\rho_k\}$ with POVM $\{\rho_k\}$, and then an observer who is not privy to the details of  the construction will see a quantum state $\rho_Q$ given by
\begin{equation}
\rho_Q = \sum_k P_k \rho_k.
\end{equation}
The Holevo information of this system is then given by
\begin{equation}
\chi(P_k, \rho_k) = S(\rho_Q)- \sum_k P_k S(\rho_k) \label{eqn:holevoOrig}
\end{equation}
where we have made clear that $\chi$ depends on the choice of POVMs. Now, considering the definition of \eqn{eqn:JHolevo} ${\cal J}(A:B)_{\Pi_k^B}$, if we choose $\rho_k \rightarrow \rho_{S|\Pi_k^B}$ then, using \eqn{eqn:conditionedoutcome}, we compute
\begin{eqnarray}
\sum_k P_k \rho_{S|\Pi_k^{\cal E}} &= & \sum_k   \Tr_{\cal E}\left[(\Pi_k^{\cal E} \otimes {\bf 1}^{S})  \rho_{\SEnv}\right]\nn
&=&\Tr_{\cal E} \rho_{\SEnv}\nn 
&=&\rho'_S,
\end{eqnarray}
where we have used the cyclic property of the trace in the first line and the POVM partition of unity $\sum_k \Pi_k^{\cal E} = {\bf 1}^{\cal E}$ in the second line. Plugging this into \eqn{eqn:holevoOrig} and comparing with \eqn{eqn:JHolevo} we get
\begin{equation}
\chi(P_k, \rho_{S|\Pi_k^{\cal E}}) = {\cal J}(S:{\cal E})_{\Pi_k^{\cal E}},
\end{equation}
and \eqn{eqn:conservation} follows. One way to interprete this quantity is the following.  The environment picks up a decoherence basis $\{\Pi_k^{\cal E}\}$, which prepares the primordial state $S$ as the ensemble $\{P_k,\rho_{S|\Pi_k^{\cal E}}\}$ with its Holevo information $\chi(P_k, \rho_{S|\Pi_k^{\cal E}})$ quantifying the \emph{maximum} amount of classical information of this ensemble (i.e the set of probabilities $\{P_k\}$) which is accessible by our measurements of the environment. From \eqn{eqn:conservation}, by minimizing discord $ \delta(S:{\cal E})_{\Pi_k^{\cal E}}$, we maximize the amount of quantum mutual information. Discord vanishes when the joint state $\rho_{\SEnv}$ is truly classical, then ${\cal I}(S:{\cal E}) \rightarrow I(S:{\cal E})$, and the Holevo bound is attained.

Finally, since the mutual information ${\cal I}(S:{\cal E})$ is independent of the probe basis $\{\Pi_k^{\cal E}\}$ while both the discord and the Holevo information is dependent on it -- this imply that the exact efficiency of the transmission is dependent on the choice of decoherence/measurement basis.  

\section{Conclusions} \label{sect:conclusions}

In this paper, we introduced quantum discord as a measure of the quantumness of cosmological correlations -- quantum discord captures deviation of classicality of the correlations beyond mere entanglement and hence is a more robust measure. We argue that even though inflation generically drives the primordial perturbations into classicality, the coupling to the environment of photons and matter fields -- which we observed as probes of the primordial perturbations -- render any observed cosmological correlations manifestly quantum.  

The environment hence serves as a quantum channel in which the information contained in the primordial perturbations is transmitted to us, and this efficiency of this transmission is dependent on the decoherence basis. Furthermore, we argue that the knowledge we gained about the primordial perturbations from an incomplete measurement of the environment, i.e. the mutual information, is bounded by the Holevo information while the ``information loss'' is exactly the quantum discord. Furthermore, since quantum discord (and the Holevo information) is dependent on measurement basis, this efficiency is deeply dependent on Nature's whim in her pick of the decoherence basis. As an illustration, we constructed a phenomenological model of a joint primordial perturbations-environment mixed state. Using this, we showed that the discord is indeed dependent on the basis of decoherence.

Since the cosmological correlations are truly quantum in nature, then they are subject to consistency tests such as Bell's or CHSH inequalities (see e.g. \cite{Campo:2005sv}). However, it is clear that even if such a test exists and is practical, we are not probing the ``primodial'' quantum nature of the fluctuations but instead we are probing the process of decoherence. While this is slightly disheartening, if one's goal is to test for the quantum origins of the primordial fluctuations, it is a test of quantum mechanics in the largest possible scale -- the scale of the cosmos itself. We leave the construction of such a statistical test to future work.

We end this work on a whimsical note -- perhaps humankind should have understood the cosmological correlations in greater detail before recklessly embarking on the campaign of  precision measurements of the CMB and hence losing possible crucial primordial information\footnote{We thank David Tong for confessing to this nightmare.}.


\acknowledgments
I would like to thank Peter Adshead, Seraphina Anderson, Daniel Baumann, Adam Brown, Aleksandra Drozd, Daniel Green, Nicholas Houston, Lam Hui, Riccardo Penco, Jens Niemeyer and David Seery for many useful conversations. I would also like to thank David Tong for conversations which motivated this work. Finally, I would like to thank my Bridgewater summer research students Toby Crisford, Cristina Cirstou and Felicity Eperon for teaching this naive cosmologist quantum information theory. This work is supported by an FQXi mini-grant and a STFC AGP grant ST/L000717/1.

\appendix
\section{Construction of Mixed Gaussian states} \label{app:purification}

Our goal is to construct a mixed joint perturbations-environment density matrix $\rho_{\SEnv}$ such that
\begin{equation}
\Tr_{\cal E} \rho_{\SEnv} = \rho_S',
\end{equation}
where $\rho_S'$ is the single mode mixed Gaussian state ansatz \eqn{eqn:ansatz}. Our strategy is to first find a Gaussian purification of $\rho_S'$, which is a two mode pure Gaussian state $\tilde{\rho}_{\SEnv}$, and then parameterize around it. It turns out that there exists a two mode Gaussian purification \cite{PhysRevA.63.032312}, whose covariance matrix $\tilde{\sigma}_{\SEnv}$ is given by
\begin{equation}
\tilde{\sigma}_{\SEnv}=\twodsquaremat{\sigma_S}{SC}{C^TS^T}{\sigma_S^{\oplus}} \label{eqn:2modeGausPure}
\end{equation}
where
\begin{equation}
C = \twodsquaremat{\sqrt{\nu^2-1}}{0}{0}{-\sqrt{\nu^2-1}},
\end{equation}
and
\begin{equation}
\sigma_S^{\oplus} = \twodsquaremat{\nu}{0}{0}{\nu}
\end{equation}
and the symplectic eigenvalue $\nu$ is the positive definite eigenvalue of the matrix $i\Omega \sigma_S$ (the symplectic matrix $\Omega$ is defined in \eqn{eqn:omega}). $S$ is the symplectic transformation matrix $S \in Sp(2,\mathbb{R})$ defined by  
\begin{equation}
S\sigma_S^{\oplus}S^T = \sigma_S~,~S\Omega S^T= \Omega.
\end{equation}
$S$ is in general not unique, but a choice of $S$ which works is the following
\begin{equation}
S=\twodsquaremat{(\lambda(\lambda+\xi))^{-1/4}}{0}{-\alpha(\lambda(\lambda+\xi))^{-1/4}}{(\lambda(\lambda+\xi))^{1/4}}.
\end{equation}
We can check that this state is pure by computing the symplectic eigenvalues of $\tilde{\sigma}_{\SEnv}$ (either by using \eqn{eqn:symplecticEV} or by calculating the positive eigenvalues of the $4\times 4$ matrix $i\Omega\tilde{\sigma}_{\SEnv}$) to find $\nu_{\pm}=1$.

To construct a mixed two mode Gaussian state, we parameterize \eqn{eqn:2modeGausPure}. There is in general a large number of ways one can make this parameterization, and the parameterization we will use, i.e. \eqn{eqn:mixedstategaussian} where 
\begin{equation}
\sigma_{\SEnv}(\beta=1,\tau=1) = \tilde{\sigma}_{\SEnv},
\end{equation} is chosen such that it illustrates the basis dependence of discord.


\bibliography{bibdiscord}


\end{document}